\documentclass{ifacconf}

\usepackage{graphicx}      
\usepackage{amsmath,amssymb} 
\usepackage{mathrsfs}  
\usepackage{natbib}        
\usepackage[hyphens]{url}

\newcommand{\proofdone}{\hspace*{\fill} \qed}

\makeatletter
\g@addto@macro\normalsize{%
  \setlength\abovedisplayskip{.3em}
  \setlength\belowdisplayskip{.3em}
  \setlength\abovedisplayshortskip{.3em}%
  \setlength\belowdisplayshortskip{.3em}%
  \setlength\parskip{.3em}
}

\usepackage{amsmath}
\usepackage{tikz}
\usepackage{xspace}

\DeclareRobustCommand{\legendline}[1]{\hspace{-3pt}
\tikz[#1,line width=0.4mm,baseline=-0.5ex]{\draw (0,0) -- (.35,0);}
\hspace{-3pt}}

\definecolor{mblue}{rgb}{0,0.4470,0.7410}
\definecolor{morange}{rgb}{0.8500,0.3250,0.0980}
\definecolor{myellow}{rgb}{0.9290,0.6940,0.1250}
\definecolor{mpurple}{rgb}{0.4940,0.1840,0.5560}
\definecolor{mgreen}{rgb}{0.4660,0.6740,0.1880}
\definecolor{mcyan}{rgb}{0.3010,0.7450,0.9330}
\definecolor{mred}{rgb}{0.6350,0.0780,0.1840}
\definecolor{mgreenblue}{rgb}{0.0,1.0,0.5}
\definecolor{parulablue}{rgb}{0.2431,0.1490,0.6588}
\definecolor{parulalblue}{RGB}{39,151,235}
\definecolor{parulagreen}{RGB}{129,204,89}
\definecolor{parulayellow}{RGB}{249,251,21}

\newcommand{\norm}[1]{\left\lVert#1\right\rVert}

\newcommand{\ltwo}{\ensuremath{\mathcal{L}_2}\xspace}

\newcommand{\dlitwo}{\ensuremath{\ell_{\mathrm{i}2}}\xspace}
\newcommand{\dltwo}{\ensuremath{\ell_2}\xspace}
\newcommand{\m}[1]{\mathcal{#1}}
\newcommand{\mr}[1]{\mathrm{#1}}

\newcommand{\hinf}{\ensuremath{\mathcal{H}_\infty}\xspace}
\newcommand{\htwo}{\ensuremath{\mathcal{H}_2}\xspace}

\DeclareMathOperator{\col}{col}

\begin{document}
\begin{frontmatter}

\title{Incremental Stability and Performance Analysis of Discrete-Time Nonlinear Systems using the LPV Framework\thanksref{footnoteinfo}} 

\thanks[footnoteinfo]{This work has received funding from the European Research Council (ERC) under the European Unions Horizon 2020 research and innovation programme (grant agreement No 714663).}

\author[TUE]{Patrick J.W. Koelewijn} 
\author[TUE,SZTAKI]{Roland T\'oth} 

\address[TUE]{Control Systems Group, Eindhoven University of Technology, P.O. Box 513, 5600 MB Eindhoven, The Netherlands, (e-mail: p.j.w.koelewijn@tue.nl, r.toth@tue.nl)}
\address[SZTAKI]{Systems and Control Laboratory, Institute for Computer Science and Control, Kende u. 13-17, H-1111 Budapest, Hungary}

\begin{abstract}                
The dissipativity framework is widely used to analyze stability and performance of nonlinear systems. By embedding nonlinear systems in an LPV representation, the convex tools of the LPV framework can be applied to nonlinear systems for convex dissipativity based analysis and controller synthesis. However, as has been shown recently in literature, naive application of these tools to nonlinear systems for analysis and controller synthesis can fail to provide the desired guarantees. Namely, only performance and stability with respect to the origin is guaranteed. In this paper, inspired by the results for continuous-time nonlinear systems, the notion of incremental dissipativity for discrete-time nonlinear systems is proposed, whereby stability and performance analysis is done between trajectories. Furthermore, it is shown how, through the use of the LPV framework, convex conditions can be obtained for incremental dissipativity analysis of discrete-time nonlinear systems. The developed concepts and tools are demonstrated by analyzing incremental dissipativity of a controlled unbalanced disk system.
\end{abstract}


\end{frontmatter}
\section{Introduction}
Stability and performance analysis are important tools to analyze quantitative properties of the behavior of a system and for the formulation of control synthesis algorithms. Many of these tools that are currently used in industry still rely on the systematic results of the \emph{Linear Time-Invariant} (LTI) framework. Most notably, the dissipativity framework introduced in \cite{Willems1972} allows for the simultaneous analysis of stability and performance of dynamical systems. These results form the cornerstone for many of the powerful and computationally efficient \emph{Linear Matrix Inequality} (LMI) based analysis and synthesis procedures that exists for LTI systems, e.g. \hinf and \htwo based analysis and control, see \cite{Scherer2015} for an overview. However, as performance demands and system complexity are ever increasing in many application fields, the ability for LTI methods to cope with these systems is getting increasingly more difficult. Hence, the use of nonlinear analysis and control methods has become of increasing interest over the last decades. Nevertheless, many of the existing nonlinear control methods only focus on ensuring stability of the closed-loop system and hence have no systematic way to incorporate performance shaping, as available in the LTI case. While some dissipativity based results for \ltwo performance and passivity analysis of nonlinear systems exist (\cite{VanderSchaft2017}), they are often cumbersome to use, requiring expert knowledge. The \emph{Linear Parameter-Varying} (LPV) framework (\cite{Shamma1988}) sought to overcome some of these issues by extending the results from the LTI framework to be used with LPV models, see \cite{Hoffmann2015} for an overview. By embedding the behavior of a nonlinear system in an LPV representation (\cite{Toth2010}), and in turn trading complexity of the problem for conservativeness of the results, the convex analysis and synthesis results to ensure stability and performance of the LPV framework could easily and systematically be applied to nonlinear systems.

However, in recent research it has been pointed out that in some cases the results of the LPV framework fail to provide the desired guarantees in order to analyze or synthesize controllers for nonlinear systems (\cite{Scorletti2015,Koelewijn2020a}). Namely, the LPV framework is only able to guarantee asymptotic stability for the origin of the nonlinear system, hence, e.g. in the case of disturbance rejection and/or reference tracking this is violated. The core issue of this is the use of the classical dissipativity framework, which expresses stability of only the origin of the system. For LTI systems, such classical dissipativity also implies stability of other forced equilibria, while for nonlinear systems this is not the case. Hence, in order to have a general stability and performance analysis framework for nonlinear systems an equilibrium independent notion of stability and dissipativity needs to be adopted.

Incremental stability (\cite{Angeli2002}), convergence (\cite{Pavlov2006}) and contraction (\cite{Lohmiller1998}) are such equilibrium independent stability notions, whereby stability of the differences between trajectories or of the variation along trajectories is considered. Incremental and differential (based on contraction) notions of dissipativity have also been considered which can be thought of as modeling the energy storage between or along trajectories analogous to the standard dissipativity framework modeling the energy storage with respect to single point of neutral storage. For \emph{Continuous-Time} (CT) nonlinear systems these results are discussed in \cite{Verhoek2020}. These methods have also been developed into convex LPV based control methods, and have successfully been applied to reference tracking and disturbance rejection of nonlinear systems (\cite{Scorletti2015,Koelewijn2019a}).

The aforementioned results on equilibrium independent stability and dissipativity analysis offer great potential to provide convex tools for nonlinear controller synthesis but are currently limited to CT nonlinear systems. Nevertheless, most control algorithms are implemented digitally, hence, analysis and control of \emph{Discrete-Time} (DT) systems plays an important role. Moreover, the recent resurgence in data-based methods for analysis and control of nonlinear systems also rely on DT systems analysis. While incremental and contraction based stability results have been extended to DT domain, see e.g. \cite{Tran2018}, similar extensions to incremental dissipativity have not yet been made to the authors' knowledge. Hence, in this paper the main contribution is to propose an extension of the CT incremental dissipativity results to DT nonlinear systems, analogous to results in \cite{Verhoek2020}, and propose LPV based convex tools to carry out the analysis.

The paper is structured as follows. In Section \ref{sec:problem}, a formal problem statement is given. In Section \ref{sec:IncrDissip}, incremental dissipativity for DT systems is discussed and as our main contribution, sufficient analysis conditions are derived to guarantee it. Section \ref{sec:lpv} gives results on how the analysis results of Section \ref{sec:IncrDissip} can efficiently be tested through the LPV framework. In Section \ref{sec:example}, as an example, the theoretical results are applied to incremental dissipativity analysis of a closed-loop discrete-time system. Finally, in Section \ref{sec:conclusion}, conclusions are drawn and future research recommendations are given.
\subsection{Notation}\vspace{-.5em}
The set of natural numbers including zero is denoted by $\mathbb{N}$. The set of real numbers is denoted by $\mathbb{R}$, where the subset $\mathbb{R}^+\subset\mathbb{R}$ corresponds to the non-negative real numbers. The set of real symmetric matrices of size $n$ by $n$ is denoted by $\mathbb{S}^n$. The space of square-summable real valued sequences $\mathbb{N}\rightarrow\mathbb{R}$ is denoted by $\ell_2$, with the norm $\norm{x}_2=\sqrt{\sum_{k=0}^\infty\norm{x(k)}^2}$, where $\norm{\cdot}$ denotes the Euclidian (vector) norm. A function $f$ is of class $\m{C}_n$, i.e. $f\in\m{C}_n$, if it is $n$-times continuously differentiable. The set of functions or sequences from $\mathbb{X}$ to $\mathbb{Y}$ is denoted by $\mathbb{Y}^\mathbb{X}$. The column vector $\begin{bmatrix}x_1^\top &\cdots &x_n^\top\end{bmatrix}^\top$ is denoted as $\col(x_1,\dots,x_n)$. The notation $A\succ 0$ ($A\succeq 0$) indicates that $A$ is positive (semi-)definite while $A\prec 0$ ($A\preceq 0$) means that $A$ is negative (semi-)definite. A function $\alpha(x)$ with $x\in X$ is positive (semi-)definite if $\alpha(x)>0$ ($\alpha(x)\geq 0$), $\forall\,x\in X\backslash\{0\}$ and $\alpha(0)=0$ and is negative (semi-)definite if $\alpha(x)<0$ ($\alpha(x)\leq 0$), $\forall\,x\in X\backslash\{0\}$ and $\alpha(0)=0$. The term that makes a matrix symmetric is denoted by $(\star)$, e.g. $(\star)^\top Qx=x^\top Qx$. 
Projection of elements or sets is denoted by $\pi_*$, where e.g. $\pi_{\mr{x,z}}(x,y,z)=(x,z)$. 
\section{Problem Statement}\label{sec:problem}
Consider a nonlinear \emph{discrete-time} (DT) dynamic system
\begin{subequations}\label{eq:nl}
\begin{align}
	x(k+1) &= f(x(k),w(k));\\
	z(k) &= h(x(k),w(k));\\
	x(0)&=x_0;
\end{align}
\end{subequations}
where $x(k) \in \m{X} \subseteq \mathbb{R}^{n_\mr{x}}$ is the state with initial condition $x_0\in\m{X}$, $w(k) \in \m{W} \subseteq \mathbb{R}^{n_\mr{w}}$ is the generalized disturbance, $z(k) \in \m{Z} \subseteq \mathbb{R}^{n_\mr{z}}$ the generalized performance and $k \in \mathbb{N}$ is the discrete-time instant. The sets $\m{X}$, $\m{W}$ and $\m{Z}$ are open and convex, containing the origin. The solutions of \eqref{eq:nl} satisfy \eqref{eq:nl} in the ordinary sense and are restricted to $k\in\mathbb{N}$. The functions $f:\m{X}\times\m{W}\rightarrow\m{X}$ and $h:\m{X}\times \m{W}\rightarrow\m{Z}$ are assumed to be Lipschitz continuous, such that $f(0,0) = 0$ and $h(0,0)=0$, and such that for all initial conditions $x_0\in\m{X}$ there is a unique solution $(x,w,z)\in(\m{X}\times\m{W}\times\m{Z})^\mathbb{N}$. We define the set of solutions of \eqref{eq:nl} as\begingroup\setlength{\jot}{-.1em}
\begin{multline}\vspace{-.5em}
	\mathfrak{B}:=\Big\lbrace (x,w,z)\in(\m{X}\times\m{W}\times\m{Z})^\mathbb{N}\mid \\(x,w,z) \text{ satisfies \eqref{eq:nl}} \Big\rbrace.
\end{multline}\endgroup
Furthermore we define the state transition map $\phi_\mr{x}:\mathbb{N}\times\mathbb{N}\times\m{X}\times \m{W}^\mathbb{N}\rightarrow\m{X}$, such that 
\begin{equation}
	x(k) = \phi_\mr{x}(k,k_0,x_0,w),
\end{equation}
which is the state $x(k)\in\m{X}$ at discrete-time instant $k\in\mathbb{N}$, with $k>k_0$, when the system is driven from $x_0\in\m{X}$ at time instant $k_0\in\mathbb{N}$ by input signal $w\in\m{W}^\mathbb{N}$.

In order to simultaneously analyze  performance and stability of nonlinear systems, dissipativity theory is widely used, which has its roots in \cite{Willems1972} for continuous-time systems and has also been extended to DT systems, see \cite{Byrnes1994}. 
\begin{defn}[Dissipativity (\cite{Byrnes1994})]\label{def:dissip}~\\
	A system of the form \eqref{eq:nl} is dissipative with respect to the supply function $s:\m{W}\times\m{Z}\rightarrow\mathbb{R}$ if there exists a positive definite storage function $V:\m{X}\rightarrow\mathbb{R}^+$ with $V(0)=0$ such that for all $k\in \mathbb{N}$ and $(x,w,z)\in\mathfrak{B}$
	\begin{equation}\label{eq:dissip}
		V(x(k+1))-V(x(k))\leq s(w(k),z(k)),
	\end{equation}
	or equivalently, for all $k\in \mathbb{N}$, $(x,w,z)\in\mathfrak{B}$ and $x_0\in\m{X}$\vspace{-.5em}
	\begin{equation}\label{eq:sumdissip}\vspace{-.5em}
		V(x(k+1))-V(x_0)\leq \sum_{j=0}^{k} s(w(j),z(j)).
	\end{equation}
\end{defn}
Performance notions such as the induced \dltwo-gain and passivity of DT nonlinear systems can be analyzed by specific choices of the supply function $s$ (\cite{VanderSchaft2017,Scherer2015}). Furthermore, under some restriction of the supply function, dissipativity implies stability of the uncontrolled system.
\begin{thm}[Stability]\label{thm:disstab}
	If a system of the form \eqref{eq:nl} is dissipative, according to Definition \ref{def:dissip}, with continuous positive definite storage function $V$ and the supply function $s$ satisfies that $s(0,z)\leq0,\,\forall\, z\in\m{Z}$ (negative semi-definite), then, the origin, i.e. $x=0$, is a stable equilibrium point of \eqref{eq:nl}. In case $s$ satisfies that $s(0,z)<0,\,\forall\, z\in\m{Z}\backslash\lbrace 0\rbrace$ (negative definite) and $s(0,0)=0$ the origin is an asymptotically stable equilibrium point.
\end{thm}
\begin{pf}
	If the system is dissipative with continuous positive definite storage function $V$ and $s(0,z)\leq0,\,\forall\, z\in\m{Z}$ it holds from \eqref{eq:dissip} that
	\begin{equation}
		V(x(k+1))-V(x(k)) \leq 0.
	\end{equation}
	Hence, the systems satisfies the condition for stability, see \cite{Kalman1960}, and $V$ is a Lyapunov function. Asymptotic stability can be proven similarly.\proofdone
\end{pf}
\begin{rem}
	The supply functions corresponding to e.g. \dltwo-gain, $s(w,z) = \gamma^2\norm{w(k)}^2-\norm{z(k)}^2$, and passivity, $s(w,z) = z(k)^\top w(k)+w(k)^\top z(k)$, satisfy the assumptions on the supply function taken in Theorem \ref{thm:disstab}.
\end{rem}

As mentioned in the introduction, the standard dissipativity framework only analyzes the internal energy of the system with respect to a single storage (equilibrium) point, often taken as the origin of the state-space associated with the nonlinear representation. However, it is often of interesest to analyze a set of equilibrium points/trajectories, e.g. in the case of reference tracking or disturbance rejection, which is cumbersome to be performed with the standard dissipativity results. Equilibrium independent dissipativity notions such as incremental dissipativity allow to efficiently handle these cases. Incremental dissipativity is an extension of the dissipativity results which takes into account multiple trajectories of a system and can be thought of as analyzing the energy flow between trajectories. The corresponding theory for CT nonlinear systems has been developed in \cite{Verhoek2020, VanderSchaft2017}. Next, we propose analogous results for incremental dissipativity of DT nonlinear systems.

\section{Incremental Stability and Performance Analysis}\label{sec:IncrDissip}
\subsection{Incremental Dissipativity}\label{sec:incrdissip}
Similar to the incremental dissipativity definition for CT systems in \cite{Verhoek2020} we define incremental dissipativity of DT nonlinear systems as follows:
\begin{defn}[Incremental Dissipativity]\label{def:incrdis}
		A system of the form \eqref{eq:nl} is incrementally dissipative with respect to the supply function $s:\m{W}\times\m{W}\times\m{Z}\times\m{Z}\rightarrow\mathbb{R}$ if there exists a storage function $V:\m{X}\times\m{X}\rightarrow\mathbb{R}^+$ with $V(x,x)=0$ such that for all $k\in \mathbb{N}$ and $(x,w,z),(\tilde{x},\tilde{w},\tilde{z})\in\mathfrak{B}$
	\begin{multline}\label{eq:incrdissip}
		V(x(k+1),\tilde{x}(k+1))-V(x(k),\tilde{x}(k))\leq \\s(w(k),\tilde{w}(k),z(k),\tilde{z}(k)),
	\end{multline}
	or equivalently, for all $k\in \mathbb{N}$, $(x,w,z),(\tilde{x},\tilde{w},\tilde{z})\in\mathfrak{B}$ and $x_0,\tilde{x}_0\in\m{X}$
	\begin{multline}\label{eq:incrdissipsum}
		V(x(k+1),\tilde{x}(k+1))-V(x_0,\tilde{x}_0)\leq \\ \sum_{j=0}^k s(w(j),\tilde{w}(j),z(j),\tilde{z}(j)).
	\end{multline} 
\end{defn}
Similar to standard dissipativity, incremental dissipativity also implies stability of the nonlinear system under some restrictions of the supply function.
\begin{thm}[Incremental stability]\label{thm:incrdisstab}
If a system of the form \eqref{eq:nl} is incrementally dissipative according to Definition \ref{def:incrdis} with a continuous storage function $V$ and the supply function $s$ satisfies that $s(w,w,z,\tilde{z})< 0,\,\forall\, w\in\m{W}$ and $\forall\,z,\tilde{z}\in\m{Z},\,z\neq\tilde{z}$ (negative definite) and $s(w,w,z,z)= 0,\,\forall\, w\in\m{W},z\in\m{Z}$, then, the system is incrementally asymptotically stable.
\end{thm}
\begin{pf}
	If $s(w,w,z,\tilde{z})< 0,\,\forall\, w\in\m{W}$ and $\forall\,z,\tilde{z}\in\m{Z},\,z\neq\tilde{z}$ and $s(w,w,z,z)\leq 0,\,\forall\, w\in\m{W},z\in\m{Z}$ 
	it holds from \eqref{eq:incrdissip} that for all $k\in\mathbb{N}$ and $x,\tilde{x}\in\pi_\mr{x}\mathfrak{B}$, $x\neq \tilde{x}$, 	\begin{equation}
		V(x(k+1),\tilde{x}(k+1))-V(x(k),\tilde{x}(k)) < 0.
	\end{equation}
	Hence, the systems satisfies the conditions for incremental asymptotic stability, see \cite{Tran2018}, and $V$ is an incremental stability Lyapunov function. Similar results implying (non-asymptotic) stability can be formulated for the case that $s(w,w,z,\tilde{z})\leq 0,\,\forall\, w\in\m{W}, z,\tilde{z}\in\m{Z},\,z\neq\tilde{z}$ (negative semi-definite), see \cite{VanderSchaft2017}.
	\proofdone
\end{pf}

In this work we will focus on supply functions of the form
\begin{equation}\label{eq:qsrsup}
	s(w,\tilde{w},z,\tilde{z})=\begin{bmatrix}
		w-\tilde{w}\\z-\tilde{z}
	\end{bmatrix}^\top 
	\begin{bmatrix}
		Q&S\\S^\top & R
	\end{bmatrix}\begin{bmatrix}
		w-\tilde{w}\\z-\tilde{z}
	\end{bmatrix},
\end{equation}
where $Q\in\mathbb{S}^{n_\mr{w}}$, $R\in\mathbb{S}^{n_\mr{z}}$ and $S\in\mathbb{R}^{n_\mr{w}\times n_\mr{z}}$. We focus on this particular family, often referred to as (incremental) (Q,S,R) supply functions, as they allow formulation of many useful performance notions, such as incremental versions of \dltwo-gain performance and passivity. Now we are ready to state our main result.

\begin{thm}[Incremental (Q,S,R)-dissipativity]\label{thm:incrdissip}
	A system of the form \eqref{eq:nl} with $f,h\in\m{C}_1$ is incrementally (Q,S,R)-dissipative, w.r.t. a supply function $s$ given by \eqref{eq:qsrsup} with $R\prec 0$ or $R=0$, if there exists a storage function 
	\begin{equation}\label{eq:incrStorage}
		V(x,\tilde{x}) = (x-\tilde{x})^\top P(x-\tilde{x}),
	\end{equation}
	with $P\succ 0$, such that for all $({x},{w})\in{\m{X}}\times{\m{W}}$
	\begin{equation}\label{eq:incrDissipFull}
	\begin{aligned}
	\begin{bmatrix}
		I & {0}\\A_\delta({x},{w}) & B_\delta({x},{w})
	\end{bmatrix}^\top \begin{bmatrix}
		-P & {0}\\{0} & P
	\end{bmatrix}\begin{bmatrix}
		I & {0}\\A_\delta({x},{w}) & B_\delta({x},{w})
	\end{bmatrix}-\\
	\begin{bmatrix}
		{0} & I\\C_\delta({x},{w})&D_\delta({x},{w})
	\end{bmatrix}^\top \begin{bmatrix}
		Q & S\\S^\top & R
	\end{bmatrix}\begin{bmatrix}
		{0} & I\\C_\delta({x},{w})&D_\delta({x},{w})
	\end{bmatrix}\preceq 0,
	\end{aligned}
\end{equation}
where
\begin{equation}\label{eq:deltaMat}
\begin{gathered}
  A_\delta({x},{w}) = \frac{\partial f}{\partial x }({x},{w}), \qquad  B_\delta({x},{w}) = \frac{\partial f}{\partial w}({x},{w}),\\
C_\delta({x},{w}) = \frac{\partial h}{\partial x }({x},{w}), \qquad  D_\delta({x},{w}) = \frac{\partial h}{\partial w}({x},{w}).
\end{gathered}
\end{equation}
\end{thm}
\begin{pf}
	According to Definition \ref{def:incrdis}, the system \eqref{eq:nl} is dissipative with respect to a supply function $s$ if \eqref{eq:incrdissip} holds for all $k\in \mathbb{N}$ and $(x,w,z),(\tilde{x},\tilde{w},\tilde{z})\in\mathfrak{B}$. Hence, \eqref{eq:nl} is incrementally (Q,S,R)-dissipative if for all $k\in \mathbb{N}$ and $(x,w,z),(\tilde{x},\tilde{w},\tilde{z})\in\mathfrak{B}$ it holds that, omitting dependence on time for brevity,
		\begin{multline}\label{eq:qsrdissiptime}
	\Delta_\mr{k}\left[ (x-\tilde{x})^\top P (x-\tilde{x})\right]- (w-\tilde{w})^\top Q(w-\tilde{w})-\\2(w-\tilde{w})^\top S(z-\tilde{z})-(z-\tilde{z})^\top R(z-\tilde{z})\leq0,
	\end{multline}
	where $\Delta_\mr{k}$ is the discrete-time difference operator, defined as $\Delta_\mr{k} v(k) = v(k+1)-v(k)$.
	For $(x,w,z),(\tilde{x},\tilde{w},\tilde{z})\in\mathfrak{B}$, define the initial conditions as $x(0):=x_0$ and $\tilde{x}(0): = \tilde{x}_0$ respectively, such that $x(k)=\phi_\mr{x}(k,0,x_0,w)$ and $\tilde{x}(k)=\phi_\mr{x}(k,0,\tilde{x}_0,\tilde{w})$. Then, define 
	\begin{gather}
		\bar{x}_0(\lambda) := \tilde{x}_0+\lambda(x_0-\tilde{x}_0),\\
		\bar{w}(k,\lambda) := \tilde{w}(k)+\lambda(w(k)-\tilde{w}(k)),\label{eq:wbar}
	\end{gather}
	with $\lambda\in[0,1]$ and
	\begin{equation}
		\bar{x}(k,\lambda):=\phi_\mr{x}(k,0,\bar{x}_0(\lambda),\bar{w}(\lambda)),
	\end{equation}
	such that $(\tilde{x}(k),\tilde{w}(k)) = (\bar{x}(k,0),\bar{w}(k,0))$ and \linebreak$(x(k),w(k)) = (\bar{x}(k,1),\bar{w}(k,1))$.
	The dynamics of $\bar{x}(\lambda)$ are then given by
\begin{subequations}\label{eq:xbardyn}
\begin{align}
	\bar{x}(k+1,\lambda) &= f(\bar{x}(k,\lambda),\bar{w}(k,\lambda));\\
	\bar{z}(k,\lambda) &= h(\bar{x}(k,\lambda),\bar{w}(k,\lambda)).
\end{align}
\end{subequations}
The first term on left hand side of inequality \eqref{eq:qsrdissiptime} can then be expressed as
\begin{equation}\label{eq:barPdiff}
	\Delta_\mr{k}\left[(\bar{x}(k,1)-\bar{x}(k,0))^\top P (\bar{x}(k,1)-\bar{x}(k,0))\right].
\end{equation}
Using the Fundamental Theorem of Calculus, \eqref{eq:barPdiff} can be expressed as
\begin{equation}
	\Delta_\mr{k}\left[\left(\int_0^1 \delta x(k,\lambda)\,d\lambda\right)^\top P \left(\int_0^1 \delta x(k,\lambda)\,d\lambda\right)\right],
\end{equation}
where $\delta x(k,\lambda) = \frac{\partial}{\partial \lambda} \bar{x}(k,\lambda)$. As $P\succ 0$, by Lemma \ref{lem:appendix}, see Appendix \ref{sec:appendix}, it holds that\vspace{-.25em}
\begin{multline}\label{eq:vdeltanormtrick}
	\Delta_\mr{k}\left[\left(\int_0^1  \delta x(k,\lambda)\,d\lambda\right)^\top P \left(\int_0^1  \delta x(k,\lambda)\,d\lambda\right)\right]\leq \\\int_0^1 \Delta_\mr{k}\left[ \delta x(k,\lambda)^\top P \delta x(k,\lambda)\right]\,d\lambda .
\end{multline}
The second term on the left-hand-side of inequality \eqref{eq:qsrdissiptime} can be expressed, using \eqref{eq:wbar}, as\vspace{-.25em}
\begin{multline}\label{eq:Qpart}
	-(\bar{w}(k,1)-\bar{w}(k,0))^\top Q(\bar{w}(k,1)-\bar{w}(k,0))=\\
	-\int_0^1 (\bar{w}(k,1)-\bar{w}(k,0))^\top Q(\bar{w}(k,1)-\bar{w}(k,0))\,d\lambda =\\\vspace{-.5em}
	-\int_0^1 \delta w(k,\lambda)^\top Q\delta w(k,\lambda)\,d\lambda,
\end{multline}
where $\delta w(k,\lambda) = \frac{\partial}{\partial \lambda} \bar{w}(k,\lambda)=w(k)-\tilde{w}(k)$ (by definition \eqref{eq:wbar}). The third term in \eqref{eq:qsrdissiptime} can similarly be expressed as\vspace{-.5em}
\begin{multline}\label{eq:Spart}\vspace{-.5em}
	-2 (\bar{w}(k,1)-\bar{w}(k,0))^\top S (\bar{z}(k,1)-\bar{z}(k,0))=\\
	-2 (\bar{w}(k,1)-\bar{w}(k,0))^\top S \int_0^1 \delta z(k,\lambda),d\lambda=\\
	-2\int_0^1 \delta w(k,\lambda)^\top S\delta z(k,\lambda)\,d\lambda,
\end{multline}
where $\delta z(k,\lambda) = \frac{\partial}{\partial \lambda} \bar{z}(k,\lambda)$. Finally, the fourth term in \eqref{eq:qsrdissiptime} can be expressed as
\begin{multline}
	-(\bar{z}(k,1)-\bar{z}(k,0))^\top R(\bar{z}(k,1)-\bar{z}(k,0))=\\
	\left(\int_0^1  \delta z(k,\lambda)\,d\lambda\right)^\top (-R) \left(\int_0^1 \delta z(k,\lambda)\,d\lambda\right).
\end{multline}
Assuming that $R\prec 0$ or $R=0$, hence, $-R\succ 0$ or $-R=0$, by Lemma \ref{lem:appendix} it holds that\vspace{-.5em}
\begin{multline}\label{eq:Rpart}\vspace{-.5em}
	\left(\int_0^1  \delta z(k,\lambda)\,d\lambda\right)^\top (-R) \left(\int_0^1 \delta z(k,\lambda)\,d\lambda\right)\leq\\
	\int_0^1 \delta z(k,\lambda) (-R)\delta z(k,\lambda)\,d\lambda.
\end{multline}
Combining the results of \eqref{eq:vdeltanormtrick}, \eqref{eq:Qpart}, \eqref{eq:Spart} and \eqref{eq:Rpart}, we obtain that, omitting dependence on time for brevity,
\begin{multline}
	\Delta_\mr{k}\left[(x-\tilde{x})^\top P (x-\tilde{x})\right]- (w-\tilde{w})^\top Q(w-\tilde{w})-\\2(w-\tilde{w})^\top S(z-\tilde{z})-(z-\tilde{z})^\top R(z-\tilde{z})\leq\\
	 \int_0^1 \Delta_\mr{k}\left[\delta x(\lambda)^\top P \delta x(\lambda)\right]-\delta w(\lambda)^\top Q\delta w(\lambda)-\\2\delta w(\lambda)^\top S\delta z(\lambda)-\delta z(\lambda) R\delta z(\lambda)\,d\lambda.
\end{multline}
Hence, if it holds that
\begin{multline}\label{eq:intdissiplm}
	\int_0^1 \Delta_\mr{k}\left[\delta x(k,\lambda)^\top P \delta x(k,\lambda)\right]-\delta w(k,\lambda)^\top Q\delta w(k,\lambda)-\\2\delta w(k,\lambda)^\top S\delta z(k,\lambda)-\delta z(k,\lambda)^\top R\delta z(k,\lambda)\,d\lambda\leq0,
\end{multline}
then, condition \eqref{eq:qsrdissiptime} holds, meaning the system is incrementally (Q,S,R)-dissipative. Furthermore, \eqref{eq:intdissiplm} holds if 
\begin{multline}\label{eq:diffdissiplm}
	\Delta_\mr{k}\left[\delta x(k,\lambda)^\top P \delta x(k,\lambda)\right]-\delta w(k,\lambda)^\top Q\delta w(k,\lambda)-\\2\delta w(k,\lambda)^\top S\delta z(k,\lambda)-\delta z(k,\lambda)^\top R\delta z(k,\lambda)\leq 0.
\end{multline}
As $f,h\in\m{C}_1$, taking the derivative w.r.t. $\lambda$ for \eqref{eq:xbardyn} results in
\begin{subequations}\label{eq:xbardyndiff}
\begin{align}
	\delta x(k+1,\lambda) &= A_\delta(\bar{x}(k,\lambda),\bar{w}(k,\lambda))\delta x(k,\lambda)+\notag\\
	&\hspace{1.3em}B_\delta(\bar{x}(k,\lambda),\bar{w}(k,\lambda))\delta w(k,\lambda);\\
	\delta z(k,\lambda) &= C_\delta(\bar{x}(k,\lambda),\bar{w}(k,\lambda))\delta x(k,\lambda)+\notag\\
	&\hspace{1.3em}D_\delta(\bar{x}(k,\lambda),\bar{w}(k,\lambda))\delta w(k,\lambda).\label{eq:zbar}
\end{align}
\end{subequations}
Hence, \eqref{eq:diffdissiplm} can be written, omitting dependence on time for brevity, as
\begin{multline}\label{eq:diffdissipineq}
	(\star)^\top P(A_\delta(\bar{x},\bar{w})\delta x+B_\delta(\bar{x},\bar{w})\delta w)-\delta x^\top P\delta x -\\ \delta w^\top Q\delta w-2\delta w^\top S(C_\delta(\bar{x},\bar{w})\delta x+D_\delta(\bar{x},\bar{w}) \delta w) -\\(\star)^\top R(C_\delta(\bar{x},\bar{w})\delta x+D_\delta(\bar{x},\bar{w}) \delta w)\leq0,
\end{multline}
which should hold for all $k\in \mathbb{N}$ and $(\bar{x},\bar{w},\bar{z})\in\mathfrak{B}$. By \cite{Willems1972}, condition \eqref{eq:diffdissipineq} can equivalently be checked by verifying \eqref{eq:diffdissipineq} on the value set, hence, checking \eqref{eq:diffdissipineq} for all $\delta x \in\mathbb{R}^{n_\mr{x}}$, $\delta w\in\mathbb{R}^{n_\mr{w}}$, $\bar{x}\in\m{X}$ and $\bar{w}\in\m{W}$ implies that \eqref{eq:diffdissipineq} holds for all $k\in \mathbb{N}$ and $(\bar{x},\bar{w},\bar{z})\in\mathfrak{B}$.
Consequently, \eqref{eq:diffdissipineq} holds if 
	\begin{multline}\label{eq:diffdissiprewritten}
		(\star)^\top \begin{bmatrix}
		-P & {0}\\{0} & P
	\end{bmatrix}\begin{bmatrix}
		I & {0}\\A_\delta({x},{w}) & B_\delta({x},{w})
	\end{bmatrix}\begin{bmatrix}
		\delta x\\\delta w
	\end{bmatrix}
	 -\\
	(\star)^\top \begin{bmatrix}
		Q & S\\S^\top & R
	\end{bmatrix}\begin{bmatrix}
		{0} & I\\C_\delta({x},{w})&D_\delta({x},{w})
	\end{bmatrix}\begin{bmatrix}
		\delta x\\\delta w
	\end{bmatrix}\leq 0,
	\end{multline}
	holds for all $\delta x \in\mathbb{R}^{n_\mr{x}}$, $\delta w\in\mathbb{R}^{n_\mr{w}}$, $x\in\m{X}$, and $w\in\m{W}$. Hence, equivalently, \eqref{eq:diffdissiprewritten} holds if for all $x,w\in\m{X}\times\m{W}$ condition \eqref{eq:incrDissipFull} holds. Consequently, if condition \eqref{eq:incrDissipFull} holds, condition \eqref{eq:qsrdissiptime} holds, which in turn implies that the system is incrementally (Q,S,R)-dissipative. \proofdone
\end{pf}
\begin{rem}\label{rem:diffstuff}
	Like in the CT case in \cite{Verhoek2020}, the DT incremental dissipativity condition derived in Theorem \ref{thm:incrdissip} can be related to differential dissipativity and contraction analysis as we will show. Namely, based on the original nonlinear system \eqref{eq:nl}, which we will refer to as the primal form of the system, with $f,h\in\m{C}_1$, we formulate the system
	\begin{equation}\label{eq:difform}
	\begin{bmatrix}
		\delta x(k+1)\\\delta z(k)
	\end{bmatrix}
	=\begin{bmatrix}
		A_\delta(x(k),w(k))&B_\delta(x(k),w(k))\\C_\delta(x(k),w(k))&D_\delta(x(k),w(k))
	\end{bmatrix}
		\begin{bmatrix}
		\delta x(k)\\\delta w(k)
	\end{bmatrix},
	\end{equation}
	where $(x,w,z)\in\mathfrak{B}$, $\delta x(k)\in\mathbb{R}^{n_\mr{x}}$, $\delta w(k)\in\mathbb{R}^{n_\mr{w}}$ and $\delta z(k)\in\mathbb{R}^{n_\mr{z}}$, often referred to as the differential form of the system, see \cite{Verhoek2020}, or variational dynamics, see \cite{Crouch1987}. It is straightforward to derive that ``standard dissipativity'', see Definition \ref{def:dissip}, of the differential form \eqref{eq:difform}, referred to as differential dissipativity, is equivalent with verifying condition \eqref{eq:incrDissipFull} in Theorem \ref{thm:incrdissip}. This is exploited in the next sections to arrive at computationally efficient checks for incremental dissipativity. See also \cite{Tran2018} and references therein for more information on differential stability and contraction analysis of DT systems.
\end{rem}


\subsection{Nonlinear Performance}
Using standard (Q,S,R)-dissipativity, many useful performance notions can be retrieved such as \dltwo-gain performance and passivity. As we will show, incremental versions of these performance notions can be introduced and analyzed using the results of Section \ref{sec:IncrDissip}.
\subsubsection{Incremental \dltwo-gain}
\begin{defn}[\dlitwo-gain]
A nonlinear system of the form \eqref{eq:nl} is said to have a finite incremental \dltwo-gain, denoted as \dlitwo-gain, if for all $w,\tilde{w}\in\ell_2$ and $x_0, \tilde{x}_0\in\m{X}$, with $(x,w,z),(\tilde{x},\tilde{w},\tilde{z}) \in\mathfrak{B}$
, there is a finite $\gamma\geq0$ and a function $\zeta(x,\tilde{x})\geq0$ with $\zeta(x,x)=0$ such that
\begin{equation}\label{eq:li2gain}
	\norm{z-\tilde{z}}_2\leq\gamma\norm{w-\tilde{w}}_2+\zeta(x_0,\tilde{x}_0).
\end{equation}
The induced \dlitwo-gain of the system is the infimum of $\gamma$ such that \eqref{eq:li2gain} still holds.
\end{defn}

Next we will show how the \dlitwo-gain of a NL system \eqref{eq:nl} can be analyzed using the results of Theorem \ref{thm:incrdissip}. 
\begin{lem}[\dlitwo-gain through incremental dissipativity] \label{lem:incrnorm}~\\ 
A nonlinear system of the form \eqref{eq:nl} has a finite \dlitwo-gain of $\gamma$ if it is incrementally (Q,S,R)-dissipative with $Q = \gamma^2 I$, $S=0$ and $R=-I$.
	\end{lem}
\begin{pf}
	If a nonlinear system of the form \eqref{eq:nl} is incrementally (Q,S,R)-dissipative with $Q = \gamma^2 I$, $S=0$ and $R=-I$, it holds that there exists a positive definite storage function $V:\m{X}\times\m{X}\rightarrow\mathbb{R}^+$ with $V(x,x)=0$ such that for all $k\in \mathbb{N}$, $(x,w,z),(\tilde{x},\tilde{w},\tilde{z}) \in\mathfrak{B}$ and $x_0,\tilde{x}_0\in\m{X}$\begingroup\setlength{\jot}{-.1em}\vspace{.2em}
	\begin{multline}
		V(x(k+1),\tilde{x}(k+1))-V(x_0,\tilde{x}_0)\leq\\ \sum_{j=0}^k \gamma^2(w(j)-\tilde{w}(j))^\top(w(j)-\tilde{w}(j))\\-(z(j)-\tilde{z}(j))^\top(z(j)-\tilde{z}(j)).\vspace{.2em}
	\end{multline}\endgroup
	If the system is incrementally (Q,S,R)-dissipative with $Q = \gamma^2 I$, $S=0$ and $R=-I$, it is also incrementally stable, as $R\prec 0$ which implies that $s(w,w,z,\tilde{z})$ is negative definite, see Theorem \ref{thm:incrdisstab}. Hence, $\lim_{k\rightarrow\infty}\norm{x(k)-\tilde{x}(k)}=0$. Therefore, $\lim_{k\rightarrow\infty}V(x(k+1),\tilde{x}(k+1))=0$, as $V(x,x)=0$, and \eqref{eq:incrdissip} becomes
	\begin{multline}
		-V(x_0,\tilde{x}_0)\leq \sum_{j=0}^\infty \gamma^2(w(j)-\tilde{w}(j))^\top(w(j)-\tilde{w}(j))\\-(z(j)-\tilde{z}(j))^\top(z(j)-\tilde{z}(j)).
	\end{multline}
	which can be written as
	\begin{equation}\label{eq:li2square}
		\norm{z-\tilde{z}}_2^2\leq \gamma^2 \norm{w-\tilde{w}}_2^2+{V(x_0,\tilde{x}_0)}.
	\end{equation}
	Hence, this implies that there exist a $\zeta(x,\tilde{x})\geq 0$ with $\zeta(x,x)=0$ such that \eqref{eq:li2gain} holds (\cite{VanderSchaft2017}).
	\proofdone
\end{pf}

\begin{thm}[\dlitwo-gain analysis]\label{thm:li2gain}
	A nonlinear system of the form \eqref{eq:nl} with $f,h\in\m{C}_1$ has a finite \dlitwo-gain of $\gamma$ if there exists a $\bar{P}\succ 0$ such that for all ${x}\in{\m{X}}$ and ${w}\in{\m{W}}$
	\begin{equation}\label{eq:matli2}
		\hspace{-.01em}\begin{bmatrix}
			\bar{P}							& A_\delta ({x},{w})\bar{P} 	& B_\delta({x},{w}) & 0\\
			\bar{P}A_\delta^\top({x},{w}) 	& \bar{P}					&0					& \bar{P}C_\delta^\top({x},{w})\\
			B_\delta^\top({x},{w})			& 0 							&\gamma I 			& D_\delta^\top({x},{w})\\
			0								& C_\delta({x},{w})\bar{P}	& D_\delta({x},{w})	&\gamma I
		\end{bmatrix}\succeq 0.\hspace{-2em}
	\end{equation} 
\end{thm}
\begin{pf}
Based on Lemma \ref{lem:incrnorm}, a nonlinear system of the form \eqref{eq:nl} has a finite \dlitwo-gain if it is incrementally (Q,S,R)-dissipative with $Q = \gamma^2 I$, $S=0$ and $R=-I$. Furthermore, based on Theorem \ref{thm:incrdissip} a nonlinear system of the form \eqref{eq:nl} with $f,h\in\m{C}_1$ is incrementally (Q,S,R)-dissipative, where $R\prec 0$ or $R=0$, with a storage function of the form \eqref{eq:incrStorage} if \eqref{eq:incrDissipFull} holds. For \dlitwo-gain analysis, $R=-I\prec 0$, hence, we can use Theorem \ref{thm:incrdissip}. Combining these results gives us that in order for \eqref{eq:nl} to have a finite \dlitwo-gain the following condition needs to be satisfied: there exists a $P\succ0$ such that for all $({x},{w})\in{\m{X}}\times{\m{W}}$
	\begin{equation}
	\begin{aligned}
	&\hspace{1em}\begin{bmatrix}
		I & 0\\A_\delta({x},{w}) & B_\delta({x},{w})
	\end{bmatrix}^\top\! \begin{bmatrix}
		-P & 0\\0 & P
	\end{bmatrix}\!\begin{bmatrix}
		I & 0\\A_\delta({x},{w}) & B_\delta({x},{w})
	\end{bmatrix}\!-\!\\
	&\begin{bmatrix}
		0 & I\\C_\delta({x},{w})&D_\delta({x},{w})
	\end{bmatrix}^\top \!\begin{bmatrix}
		\gamma^2I & 0\\0 & -I
	\end{bmatrix}\!\begin{bmatrix}
		0 & I\\C_\delta({x},{w})&D_\delta({x},{w})
	\end{bmatrix}\!\preceq \!0.
	\end{aligned}
	\end{equation}
This condition can simply be rewritten into \eqref{eq:matli2} by defining $\bar{P} = \gamma P^{-1}$, taking a Schur complement and applying a congruence transformation. \proofdone
%
\end{pf}

\subsubsection{Incremental passivity}
Similar to the definitions in \cite{VanderSchaft2017,Verhoek2020} we define (DT) incremental passivity as follows:
\begin{defn}[Incremental passivity]\label{def:incrpass}
	A nonlinear system of the form \eqref{eq:nl} is said to be incrementally passive if it is incrementally dissipative, see Definition \ref{def:incrdis}, with respect to the supply function 
	\begin{equation}\label{eq:incrpasssup}
		s(w,\tilde{w},z,\tilde{z}) = (w-\tilde{w})^\top(z-\tilde{z})+(z-\tilde{z})^\top(w-\tilde{w}).
	\end{equation}
\end{defn}
\begin{thm}[Incremental passivity analysis]
	A nonlinear system of the form \eqref{eq:nl} with $f,h\in\m{C}_1$ is incrementally passive if there exists a $P\succ 0$ such that for all ${x}\in{\m{X}}$ and ${w}\in{\m{W}}$
	\begin{equation}\label{eq:incrpassmatfull}
		\hspace{-.1em}\begin{bmatrix}
			P&A^\top_\delta({x},{w}) P & C^\top_\delta({x},{w}) \\
			PA_\delta({x},{w})& P&PB_\delta({x},{w})\\
			C_\delta({x},{w})&B^\top_\delta({x},{w})P&D_\delta({x},{w})+D^\top_\delta({x},{w})
		\end{bmatrix}\succeq 0.\hspace{-2em}
	\end{equation}
\end{thm}
\begin{pf}
	According to Definition \ref{def:incrpass}, a system of the form \eqref{eq:nl} is incrementally passive if it is incrementally dissipative with respect to the supply function $s$ given by \eqref{eq:incrpasssup}. This supply function can also be written in (Q,S,R) form, see \eqref{eq:qsrsup}, by taking $Q=0$, $S=-I$ and $R=0$. By using the results of Theorem \ref{thm:incrdissip}, and filling in $Q=0$, $S=-I$ and $R=0$ in condition \eqref{eq:incrDissipFull}, it  can simply be rewritten into \eqref{eq:incrpassmatfull} by taking a Schur complement and congruence transformation.\proofdone
\end{pf}
\begin{rem}
	Note that the obtained conditions for \dlitwo-gain and incremental passivity analysis result in checking positive semi-definiteness of a matrix, while in literature these, or similar conditions, are often found as positive definiteness checks. The positive definite versions of the conditions can simply retrieved by making the incremental dissipativity check strict, i.e. changing $\leq$ to $<$ in \eqref{eq:incrdissip}, which then imply the strict versions of the conditions found in this paper.
	\end{rem}

\section{Convex Analysis using the LPV Framework}\label{sec:lpv}
As shown in Section \ref{sec:IncrDissip}, the condition for incremental dissipativity can be written in terms of LMIs which needs to be checked for infinitely many pairs $({x},{w})\in\m{X}\times\m{W}$. This is similar to the problem for performance and stability analysis of LPV systems where LMIs need to be checked for infinitely many values of a scheduling-variable $\rho\in \m{P}\subset\mathbb{R}^{n_\rho}$. Hence, we make use of the developed LPV approaches to make the proposed incremental dissipativity conditions computationally feasible. 

As we have shown in Section \ref{sec:IncrDissip} the resulting incremental dissipativity conditions for a system \eqref{eq:nl} are related to standard dissipativity of its differential form \eqref{eq:difform}. Hence, we embed the differential form of the nonlinear system in an LPV model.
\begin{defn}[LPV embedding]\label{def:lpvemb}
	Assume we have a nonlinear system of the form \eqref{eq:nl} with $f,h\in\m{C}_1$ and with differential form given by \eqref{eq:difform}. The LPV state-space model given by
		\begin{equation}\label{eq:lpvdiff}
	\begin{bmatrix}
		\delta x(k+1)\\\delta z(k)
	\end{bmatrix}
	=\begin{bmatrix}
		A(\rho(k))&B(\rho(k))\\C(\rho(k))&D(\rho(k))
	\end{bmatrix}
		\begin{bmatrix}
		\delta x(k)\\\delta w(k)
	\end{bmatrix},
	\end{equation}
	where $\rho(k)\in\m{P}\subset\mathbb{R}^{n_\rho}$ is the scheduling-variable is an LPV embedding on the compact convex region $\mathscr{X}\times\mathscr{W}$ of the differential form \eqref{eq:difform} if there exists a function, called the scheduling-map, $\psi:\mathbb{R}^{n_\mr{x}}\times\mathbb{R}^{n_\mr{w}}\rightarrow\mathbb{R}^{n_\rho}$ such that under a given choice of function class for $A,\,\dots,\,D$, e.g. affine, polynomial, etc., $A(\psi(x,w))=A_\delta(x,w)$, $\dots$, $D(\psi(x,w))=D_\delta(x,w)$ for all $x\in\mathscr{X}$, $w\in\mathscr{W}$ and $\psi(\mathscr{X},\mathscr{W})\subseteq \m{P}$ where $\m{P}$ is a (minimal) convex hull with $n$ vertices. By specific choice of the embedding region $\mathscr{X}\times\mathscr{W}$, either the full state-space can be embedded of the original NL model \eqref{eq:nl} in which case $\mathscr{X}\times\mathscr{W}\supseteq\m{X}\times\m{W}$ or part of the state-space can be embedded, in which case $\mathscr{X}\times\mathscr{W}\subseteq\m{X}\times\m{W}$.
	\end{defn}

\begin{thm}[Incremental Dissipativity LPV Analysis]\label{thm:incrdissipLPV}	~\\
	Assume a system of the form \eqref{eq:nl} with $f,h\in\m{C}_1$ and with an LPV embedding on the compact region $\mathscr{X}\times\mathscr{W}$ of its differential form given by \eqref{eq:lpvdiff}, see Definition \ref{def:lpvemb}, with scheduling-variable $\rho$, scheduling-map $\psi$ and such that $\psi(\mathscr{X}\times\mathscr{W})\subseteq\m{P}$. The system \eqref{eq:nl} is incrementally (Q,S,R)-dissipative, on the region $\mathscr{X}\times\mathscr{W}$ with respect to the supply function $s$, given by \eqref{eq:qsrsup} with $R\prec0$ or $R=0$, and with storage function $V$ given by \eqref{eq:incrStorage}	with $P\succ 0$, if for all $\rho\in\m{P}$ 
	\begin{equation}\label{eq:incrDissipFullLPV}
	\begin{aligned}
	\begin{bmatrix}
		I & 0\\A(\rho) & B(\rho)
	\end{bmatrix}^\top \begin{bmatrix}
		-P & 0\\0 & P
	\end{bmatrix}\begin{bmatrix}
		I & 0\\A(\rho) & B(\rho)
	\end{bmatrix}-\\
	\begin{bmatrix}
		0 & I\\C(\rho)&D(\rho)
	\end{bmatrix}^\top \begin{bmatrix}
		Q & S\\S^\top & R
	\end{bmatrix}\begin{bmatrix}
		0 & I\\C(\rho)&D(\rho)
	\end{bmatrix}\preceq 0.
	\end{aligned}
\end{equation}
\end{thm}
\begin{pf}
	A system of the form \eqref{eq:nl} with $f,h\in\m{C}_1$ is incrementally (Q,S,R)-dissipative on the region $\mathscr{X}\times\mathscr{W}$, there exists a $P\succ 0$ such that, for all $({x},{w})\in\mathscr{X}\times \mathscr{W}$, condition \eqref{eq:incrDissipFull} holds. As $\rho\in\m{P}$ and $\psi(\mathscr{X},\mathscr{W})\subseteq\m{P}$, checking whether there exists a $P\succ 0$ such that for all $\rho\in\m{P}$ condition \eqref{eq:incrDissipFullLPV} holds implies that for all $({x},{w})\in\mathscr{X}\times \mathscr{W}$ condition \eqref{eq:incrDissipFull} holds.

\end{pf}
The resulting condition that needs to be checked for incremental (Q,S,R)-dissipativity using the LPV framework in Theorem \ref{thm:incrdissipLPV} is similar to the condition that needs to be checked for standard (Q,S,R)-dissipativity of DT LPV systems, see e.g. the \dltwo-gain results in \cite{Apkarian1995,DeOliveira2002}. Note however the proposed incremental dissipativity analysis uses an LPV embedding of the differential form \eqref{eq:difform}, while standard dissipativity analysis uses an LPV embedding of the primal form \eqref{eq:nl}. As the proposed analysis results for incremental (Q,S,R)-dissipativity can be casted a standard (Q,S,R)-dissipativity analysis problem of an LPV system, all of the techniques to reduce the evaluation of an infinite set of LMIs to only checking a finite set of LMIs from the LPV framework can be used. Often for this, $A,\,\dots,\,D$ are needed to be restricted to an affine function in the embedding \eqref{eq:lpvdiff}. The most common techniques are polytopic, multiplier or gridding-based approaches, see \cite{Hoffmann2015} for an overview. Although the same tools from the LPV framework can be used for checking incremental dissipativity and `standard' dissipativity of nonlinear systems, we would like to stress that the underlying dissipativity and stability concepts are very different. Namely, using the incremental dissipativity tools developed in this paper, global stability and performance guarantees can be given for the nonlinear system, while standard dissipativity tools can only provide performance and stability analysis with respect to single equilibrium point, often the origin of the state-space representation of the nonlinear system.

\section{Example}\label{sec:example}
\begin{figure}
	\centering
	\includegraphics[scale=0.2]{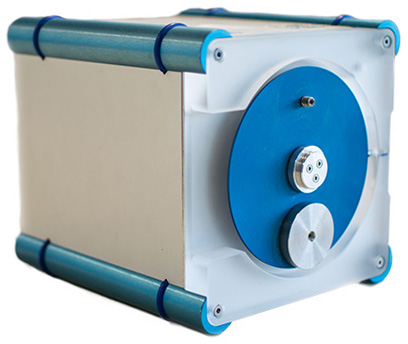}\vspace{-.5em}
	\caption{Unbalanced disk setup.}
	\label{fig:ubdisk}
\end{figure}

In this section, we apply the results from Section \ref{sec:IncrDissip} in order to analyze incremental dissipativity of a controlled unbalanced disk. The CT dynamics of the unbalanced disk system, see Fig. \ref{fig:ubdisk}, can be expressed in nonlinear state space form by neglecting the fast electrical dynamics: 
\begin{subequations}\label{eq:ubdisk}
\begin{align}
	\dot{x}_1(t) &= x_2(t);\\
	\dot{x}_2(t) &= \frac{Mgl}{J}\sin(x_1(t))-\frac{1}{\tau }x_2(t)+\frac{K_m}{\tau}u(t);
	\end{align}
\end{subequations}
where $M$ is the mass attached to the disk and $x_1$ [rad] its angular position, $x_2$ [rad/s] its angular velocity, $u$ [V] is the control input voltage, $g$ is the gravitational acceleration, $l$ the length of the pendulum, $J$ the inertia of the disk and $K_m$ and $\tau$ are the motor constant and friction coefficient respectively. The values of the physical parameters of the system are given in Table \ref{tb:param}.

\begin{figure}
	\centering
	\includegraphics[scale=1]{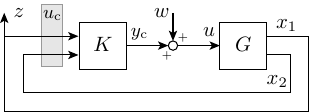}
	\caption{Closed-loop interconnection of DT controller $K$ and discretized dynamics of the unbalanced disk $G$.}
	\label{fig:clp}
\end{figure}

We discretize equation \eqref{eq:ubdisk} using a fourth order Runge-Kutta (RK4) method, where the control input is assumed to be constant over the sampling period. More specifically, assuming the CT dynamics are $\dot{x}(t) = f_\mr{c}(x(t),u(t))$, we have the RK4 discretized dynamics given by
\begin{equation}
	x(k+1) = x(k)+\frac{T_\mr{s}}{6}(\varphi_1(k)+2\varphi_2(k)+2\varphi_3(k)+\varphi_4(k)),
\end{equation}
where
\begin{subequations}
\begin{align}
	\varphi_1(k) &= f_\mr{c}(x(k),u(k)),\\
	\varphi_2(k) &= f_\mr{c}\left(x(k)+\tfrac{T_\mr{s}}{2}\varphi_1(k),u(k)\right),\\
	\varphi_3(k) &= f_\mr{c}\left(x(k)+\tfrac{T_\mr{s}}{2}\varphi_2(k),u(k)\right),\\
	\varphi_4(k) &= f_\mr{c}\left(x(k)+T_\mr{s}\varphi_3(k),u(k)\right),
\end{align}
and where $T_\mr{s}$ is the sample time.
\end{subequations}
Applying this method to the CT dynamics of the unbalanced disk \eqref{eq:ubdisk}, with a sample time $T_\mr{s}=\tfrac{1}{20}$ second, results in a DT nonlinear state-space representation of the form
\begin{equation}\label{eq:dtubdisk}
	x(k+1) = f(x(k),u(k)),
\end{equation}
where $x(k) = \col(x_1(k),x_2(k))$.
For the discretized version of the unbalanced disk \eqref{eq:dtubdisk}, a DT LTI controller is heuristically designed in order to achieve reference tracking. This controller is given by
\begin{subequations}\label{eq:dtcontroller}
	\begin{align}
		x_\mr{c}(k+1) &= x_\mr{c}(k)+B_\mr{c} u_\mr{c}(k);\\
		y_\mr{c}(k) &= C_\mr{c}x_\mr{c}(k)+D_\mr{c} u_\mr{c}(k);
	\end{align}
\end{subequations}
where $x_\mr{c}$ is the state, $u_\mr{c}$ is the input and $y_\mr{c}$ is the output of the controller. For the LTI controller, $B_\mr{c} = \begin{bmatrix}1 & 0\end{bmatrix}$, $C_\mr{c} = -0.5$ and $D_\mr{c} = \begin{bmatrix}-10&-1\end{bmatrix}$ are chosen, corresponding to a PID controller. The closed-loop interconnection of plant and controller is given in Fig. \ref{fig:clp}, where $K$ is the DT LTI controller \eqref{eq:dtcontroller}, $G$ is the discretized unbalanced disk dynamics \eqref{eq:dtubdisk}, $w$ is the input disturbance and $z$ the angle of the disk. The controller $K$ in this configuration can be thought of as a PID controller for regulation of the disk angle at zero and rejection of constant input disturbances. The closed-loop interconnection results in a system of the form \eqref{eq:nl}.
\begin{table}
\begin{center}
\caption{Parameters of the unbalanced disk.}\label{tb:param}\vspace{-.7em}
\begin{tabular}{cccccc}
$M$ 	& $g$ & $l$		& $J$ 					& $K_m$ & $\tau$  \\\hline
0.076 	& 9.8 & 0.041 	& 2.4$\cdot 10^{-4}$	 	& 11 	& 0.40 \\
\end{tabular}
\end{center}\vspace{.5em}
\end{table}

Using Definition \ref{def:lpvemb}, the differential form of the closed-loop dynamics of the DT LTI controller and discretized unbalanced disk dynamics is computed and is embedded in an LPV representation on the compact region\footnote{One can also restrict the controller state $x_\mr{c}$ and the generalized disturbance $w$ to compact sets such that $u(k)\in[-10,10]$, although these are not explicitly given.} $x_1(k)\in[-\pi,\pi]$, $x_2(k)\in[-10,10]$ and $u(k)\in[-10,10]$, with scheduling-variable $\rho=\col(\rho_1,\rho_2,\rho_3)=\col(x_1,x_2,u)$. Next, an upper-bound for the induced \dlitwo-gain of the closed-loop interconnection on the compact region is computed using the results of Theorem \ref{thm:li2gain} and Theorem \ref{thm:incrdissipLPV}. To reduce the infinite set of LMIs to a finite set of LMIs, a gridding-based method is used, due to the complexity of the discretized plant, whereby the compact region of the LPV embedding is equidistantly gridded with 11 grid-points in each dimension resulting in a total of 1331 grid-points. 	Solving the optimization problem results in an upper-bound for the induced \dlitwo-gain of $\gamma = 0.220$ for the closed-loop interconnection on the compact region. In order to compute the closed-loop \dltwo-gain of the closed-loop interconnection, the DT primal form in the plant \eqref{eq:dtubdisk} is embedded in a grid-based LPV model using the technique described in \cite{Koelewijn2021} on the aforementioned equidistant grid. The closed-loop interconnection of the LTI controller and primal form of the plant obtains an upper-bound for the \dltwo-gain\footnote{Note that the \dltwo-gain is smaller than the \dlitwo-gain, as the \dlitwo-gain is a stronger notion.} of $\gamma_{\ell_2}=0.219$

For comparison, an LPV version of the controller is also heuristically designed, where $B_\mr{c}$ is taken the same as for the LTI controller \eqref{eq:dtcontroller}, but $C_\mr{c}$ and $D_\mr{c}$ are made parameter-varying by taking $C_\mr{c}(\rho) = -0.5-\tfrac{1}{20}\sin(\rho_1)$ and $D_\mr{c}(\rho) = \begin{bmatrix}
	-10-2\cos(\rho_1)&-1
\end{bmatrix}$ (hence, they only vary in $\rho_1=x_1$). For the closed-loop interconnection of the LPV controller and the primal form of the plant an upper-bound for the \dltwo-gain is computed using a standard grid-based LPV method, resulting in $\gamma_{\ell_2} = 0.179$, which is better than the closed-loop interconnection with the LTI controller. However, unlike the closed-loop with the LTI controller, the closed-loop with LPV controller does \emph{not} have a bounded \dlitwo-gain.

\begin{figure}
	\centering
	\vspace{-1em}
	\includegraphics[scale=1]{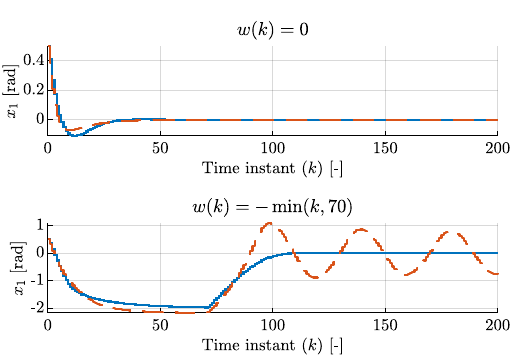}
	\caption{Angle of the disk in closed-loop with the LTI controller (\legendline{mblue}) and LPV controller (\legendline{morange,dashed}) for different input disturbances $w$.}
	\label{fig:stateresp}
\end{figure}

In Fig. \ref{fig:stateresp}, simulation results of trajectory of the angle of the disk for both the interconnection of the discrete-time plant with the LTI controller and with the LPV controller for different input disturbances $w$ are shown\footnote{Note that during simulation all the scheduling-variables stayed within the compact-set.}. In the case that $w(k)=0$, the closed-loop with the LPV controller has a faster response and less overshoot compared to the closed-loop with LTI controller. This is also to be expected, as the closed-loop $\dltwo$-gain with LPV controller ($\gamma_{\ell_2}=0.179$) is lower than that of the closed-loop system with LTI controller ($\gamma_{\ell_2}=0.219$). However, in the case that $w(k)=-\min(k,70)$, it can be seen that while the LTI controller is still able to reject the disturbance when it becomes constant (at $k=70$), the closed-loop with the LPV controller ends up in a limit cycle and is not able to reject the disturbance as it is not incrementally dissipative. This behavior is similar to what is seen in the continuous-time case (\cite{Koelewijn2020a}). This highlights the importance of analyzing stability and performance of nonlinear systems using incremental dissipativity instead using only standard dissipativity based notions to fully exploit the potential of controller synthesis methods, such as LPV synthesis, for forced equilibrium stabilization and tracking control of nonlinear systems.


\section{Conclusion}\label{sec:conclusion}
In this paper extensions of the CT incremental dissipativity framework to DT nonlinear systems have been proposed, along with convex conditions to analyze it. The proposed analysis condition use the LPV framework for efficient computation of the various incremental performance notions. The DT incremental (Q,S,R)-dissipativity results, analogous to the CT results, show that incremental (Q,S,R)-dissipativity of DT systems can be evaluated by evaluating `standard' dissipativity of their differential form, i.e. the dynamics of the variations along the systems trajectory. Moreover, using the LPV framework, this problem can then be casted as a standard dissipativity check of an LPV model, which allows for the many computational techniques of the LPV framework to be used to efficiently solve nonlinear performance analysis problems using convex optimization. These results pave the way for development of efficient synthesis techniques to ensure incremental dissipativity of DT nonlinear systems. For future research, we aim to develop such synthesis techniques and extend the analysis results to allow for a state dependent quadratic matrix of the storage function in order to reduce conservativeness.
\bibliography{Ref}

\appendix
\section{Norm Integral Inequality}\label{sec:appendix}
\begin{lem}\label{lem:appendix}
	Given a positive definite $M\in\mathbb{R}^{n\times n}$, i.e. $M\succ 0$, and a continuous function $\phi:[0,1]\rightarrow\mathbb{R}^n$, then 
	\begin{equation}\label{eq:apptoproof}\vspace{-.5em}
		\left(\int_0^1\phi(t)\,dt\right)^\top M \left(\int_0^1\phi(t)\,dt\right)\leq \int_0^1 \phi(t)^\top M\phi(t)\,dt 
	\end{equation}
\end{lem}
\begin{pf}
	As $M$ is positive, we can define the Euclidean vector space with\vspace{-.5em}
	\begin{equation}\label{eq:normdef}
		\norm{v} := \sqrt{v^\top M v},
	\end{equation}
	where $v\in\mathbb{R}^n$.
By the Cauchy-Schwarz inequality, for a continuous function $\phi:[0,1]\rightarrow\mathbb{R}^n$
\begin{equation}\label{eq:A3}
	\norm{\int_0^1 \phi(t) \,dt}\leq \int_0^1\norm{\phi(t)}\,dt,
\end{equation}
see \cite{Rudin1976}.
Furthermore, it also holds that for a function $\psi:[0,1]\rightarrow\mathbb{R}$
\begin{multline}\label{eq:A4}
	\left\vert \int_0^1 \psi(t)\,dt\right\vert^2\leq \left(\int_0^1 1\,dt\right)\left(\int_0^1 \vert \psi(t)\vert^2 \,dt\right) =\\ \left(\int_0^1 \vert \psi(t)\vert^2\,dt\right).
\end{multline}
Hence, using \eqref{eq:A3} and \eqref{eq:A4}, with $\psi(t)=\norm{\phi(t)}$, we get
\begin{equation}
	\norm{\int_0^1 \phi(t) \,dt}^2\leq \left(\int_0^1\norm{\phi(t)}\,dt\right)^2\leq \int_0^1\norm{\phi(t)}^2\,dt.
\end{equation}
Using the norm definition \eqref{eq:normdef}, this results in \eqref{eq:apptoproof}.
\end{pf}

\end{document}